# Non-adiabatic Semiclassical Dressed States

I. G. Koprinkov
Department of Applied Physics, Technical University of Sofia, 1756 Sofia, Bulgaria
E-mail:igk@vmei.acad.bg

*Abstract*

We introduce non-adiabatic semiclassical dressed states for a quantum system interacting with an electromagnetic field of variable amplitude and phase, and presence of dumping. We also introduce a generalized adiabatic condition, which allows finding of closed form solution for the dressed states. The influence of the non-adiabatic factors on the dressed states due to the amplitude and phase field variations and dumping has been found.

**1. Introduction**

The interaction is fundamental phenomenon, able to reveal the physical features of the interacting systems, in which they participate in an equivalent and symmetric way. Although the physical results must be quantitatively independent on the physical basis, the physical picture may look qualitatively different within different basis states. In principle, the most adequate physical description must as close as possible simulate the physical reality. In that relation, the dressed states (DSs) [1, 2], which represent a sum of products between the corresponding states of the quantum system (QS) and the quantized electromagnetic field (EMF), naturally correspond to the above mentioned equivalence and symmetry in the interaction. The corresponding semiclassical DSs, frequently called adiabatic states, also possess that property.

The semiclassical DSs are derived either by solving the corresponding equations of motion for the time-dependent state amplitudes [3-5], or by finding suitable unitary transformation that diagonalizes the respective Hamiltonian [6-8]. To solve the problem, one invokes the adiabatic approximation. In the usual application of the adiabatic approximation one neglects the terms associated with the field variations (time derivatives) [5], considered as non-adiabatic factors. Another source of non-adiabaticity arises from dumping [5-7]. Although the early treatment of dumping concerns the wave function [9], now it is typically performed within the density matrix approach. Thus, the DSs derived so far do not actually incorporate non-adiabatic factors coming from the filed and dumping.

The concept of adiabatic evolution has basic importance because it becomes related with other basic subjects: the quantum adiabatic theorem [10], adiabatic transitions [11, 12], adiabatic separability between spatial and temporal variables (a generalized adiabatic concept) [13], the original introduction of Barry phase [14], etc. Higher order adiabatic evolution has been also studied [13, 15, 16]. In the semiclassical limit ($\hbar \to 0$), the amplitude of the non-adiabatic transition in the area of complex crossing between two adiabatic potential surfaces can be found within the adiabatic evolution under a time dependent Hamiltonian without reference to the non-adiabatic coupling, responsible for the transition [17-19]. Non-adiabatic effects due to rapidly fluctuating fields, treated by stochastic methods, were subject of a number of studies [20-22]. As the experiments show [6], even at well-satisfied adiabatic condition, the transfer of population between different DSs is not negligible. Some times, it is hardly to distinguish in practice the adiabatic from the non-adiabatic contribution. The non-adiabatic effects depend, in general, on the non-adiabatic factors and to well understand the non-adiabatic dynamic of the

QS, the non-adiabatic factors must be well specified and traced. In addition, study of the non-adiabatic effects may reveal some particular features of the field-matter interaction.

In the present work we propose an approach that allows finding of nonperturbative closed form solution for the semiclassical DSs, taking into account non-adiabatic contributions from both, the field and the dumping. To trace the non-adiabatic effects, the non-adiabatic factors were explicitly introduced in the initial quantum equations of motion. The filed is considered non-adiabatically varying but not fluctuating. The dumping, which also causes non-adiabatic effects [5-7], is described, as usual, by phenomenological dumping term. The main difficulties, however, arise from the field non-adiabaticity, which leads to equations that, in general, are not solvable analytically. The problem is treated dynamically and no stochastic methods have been used. This allows tracing the pure non-adiabatic dynamic of the QS. We introduce generalized adiabatic condition, which helps solving the problem keeping the leading order of non-adiabaticity. The field non-adiabaticity should be considered weak within the generalized adiabatic condition. In contrast to the usual semiclassical DSs, the DSs so obtained will be considered as non-adiabatic DSs. To the best of our knowledge, non-adiabatic DSs are introduced for the first time. They allow studying some general properties of the non-adiabatic processes by means of closed form expressions. While not generally separated, the amplitude and the phase filed variations were found to cause specific effects on the parameters of the DSs. Some physical aspects in the formation of the real and virtual components of the DSs have been discussed. The interaction between a QS and an EMF, when the later is near resonant with some two-levels of the QS, represents basic phenomenon for many two-level or multilevel problems. In this way, the present results become closely related with some rapidly developing fields of research, as, e.g., the manipulation and coherent control of quantum processes [23-25], laser cooling [26], coherent population trapping [27], quantum information [28, 29], etc. Within the quantum information problem, the two-level QS represents single quantum bit (qubit) and the solution found here can be also considered as an analytic form of single qubit internal dynamic influenced by non-adiabatic factors. Because the near-resonant excitation takes place in many practical cases, the non-adiabatic effects must be expected and taken into account from both, the qualitative and quantitative reasons.

## 2. Derivation of the Non-adiabatic Dressed States

Consider the interaction between a non-degenerate two-level QS, having an electric dipole allowed transition, and a near resonant linearly polarized EMF, (Fig.1).

$$E = (1/2) E_o [\exp(i\Phi) + c.c.] , \qquad (1)$$

where $E_o(t)$ and $\Phi(t) = \omega t + \varphi(t)$ are field amplitude and phase, respectively. The relaxation processes due to interaction of the QS with the macroscopic environment (other QSs from the ensemble and zero-point vacuum fluctuations) are also considered and their entire effect will be described by a complex dumping rate $\gamma = \gamma' - i\gamma''$. The amplitude $E_o^{-1} \partial_t E_o$ and phase $\partial_t \varphi$ field variations ($\partial_t^n \equiv \partial^n/\partial t^n$), and the dumping $\gamma$ are considered as non-adiabatic factors acting on the QS. The Hamiltonian of the QS under consideration (within the bare states representation, $\hat{H}_o |j\rangle = \hbar \omega_j |j\rangle$, $j = 1,2$) is

$$\hat{H} = \sum_j \hbar \omega_j |j\rangle\langle j| - \mu E (|1\rangle\langle 2| + h.c.) - i\hbar \frac{\gamma}{2} |2\rangle\langle 2| \qquad , \qquad (2)$$

where $|j\rangle$ are the ground ($j=1$) and excited ($j=2$) bare states, $\omega_j$ are their eigenfrequencies, and $\mu = \langle 1|\hat{\mu}|2\rangle = \langle 2|\hat{\mu}|1\rangle$ is the dipole matrix element.

When the EMF is switched-on, the QS will, in general, be in a superposition (coherent) state $|\Psi(\vec{r},t)\rangle$, which obey the time-dependent Schrödinger's equation $\hat{H}|\Psi(\vec{r},t)\rangle = i\hbar\partial_t|\Psi(\vec{r},t)\rangle$. The superposition state, initially expressed within the bare state basis, is

$$|\Psi(\vec{r},t)\rangle = \sum_j a_j(t)|j(\vec{r})\rangle\exp(-i\omega_j t) \quad , \qquad (3)$$

where $a_j(t)$ are time-dependent amplitudes.

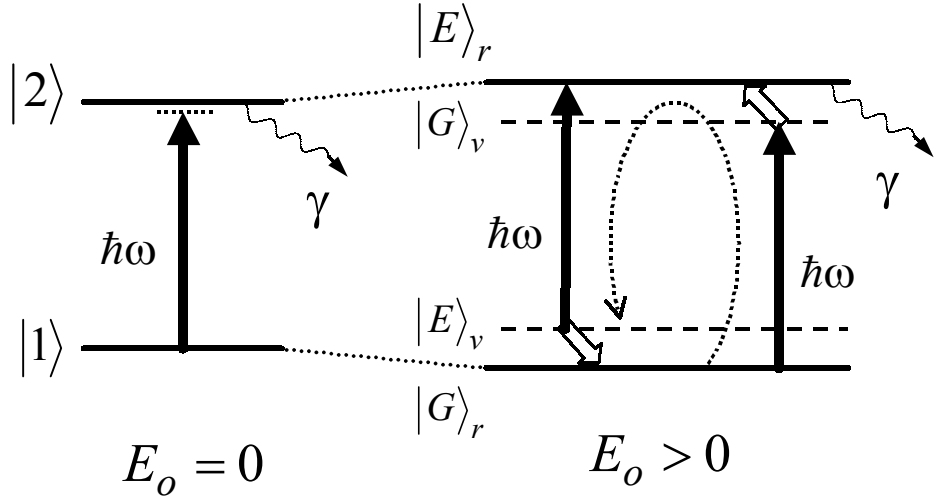

Fig.1: Bare ($E_o=0$) and adiabatic ($E_o>0$) states of a quantum system. The bold arrows show the optical pumping, the empty arrows show the non-adiabatic processes, and the wavy arrows show the dumping.

We derive the DSs from the equations of motion of the time-dependent amplitudes. Following the standard procedure and eliminating the anti-resonant terms, rotating-wave approximation (RWA), the equations of motion are:

$$\begin{aligned}\partial_t a_1(t) &= (i/2)\Omega(t)\exp[-i\Delta\Phi(t)]\, a_2(t) \\ \partial_t a_2(t) &= -(1/2)\gamma\, a_2(t) + (i/2)\Omega(t)\exp[i\Delta\Phi(t)]\, a_1(t)\end{aligned} \qquad (4)$$

where $\Omega(t)=\mu E_o(t)/\hbar$ is the on-resonance Rabi frequency, $\Delta\Phi = \Delta\omega\, t - \varphi(t)$, and $\Delta\omega = \omega_2 - \omega_1 - \omega$ is the zero-field frequency detuning. Eliminating $a_2(t)$ from Eqs.(4) and making use of substitution

$$a_1(t) = f(t)\exp\left\{-(i/2)\int_0^t \Delta\tilde{\omega}'dt'\right\} \quad , \qquad (5)$$

one obtains the "normal" form equation

$$\partial_t^2 f + (1/4)\tilde{\Omega}'^2\, f = 0 \quad , \qquad (6)$$

where $\tilde{\Omega}'=[\Omega^2 + \Delta\tilde{\omega}'^2 - i2\partial_t\Delta\tilde{\omega}']^{1/2}$ will be called instantaneous off-resonance Rabi frequency and $\Delta\tilde{\omega}' = \Delta\omega - \partial_t\varphi - \gamma''/2 - i(\gamma'/2 - \Omega^{-1}\partial_t\Omega)$ has meaning of instantaneous complex frequency detuning.

In general, Eq.(6) does not have exact analytical solution. That is why we will look for approximate solution adapting an approach of Ref. [30]. Assume that $f(t)$ has the form $f(t) = T(t)R[S(t)]$, Eq. (6) becomes

$$T(\partial_t S)^2 \partial_s^2 R + (2\,\partial_t T\,\partial_t S + T\partial_t^2 S)\partial_s R + (\partial_t^2 T + T\tilde{\Omega}'^2/4)R = 0 \qquad (7)$$

where the new functions $T$ and $R$, and the new variable $S(t)$ will be determined so as to make Eq. (7) solvable. As can be shown, the second term in Eq. (8) can be eliminated if $T(t)$ has the form $T = C(\partial_t S)^{-1/2}$, where $C$ is an arbitrary constant that, without loss of generality, can be chosen $C = 1$. This reduces Eq. (7) to

$$\partial_s^2 R + (\partial_t S)^{-2}(\partial_t^2 T/T + \tilde{\Omega}'^2/4)R = 0 \qquad (8)$$

Within the semiclassical consideration [30], the first term in the brackets in Eq. (8) is ignored in comparison with the second one due to the large factor $1/\hbar$ existing there. Such factor would also formally appear here if in Eq. (3) we introduce energies of the bare states, instead of frequencies. Instead of such an approach, the solution of the problem will be done here introducing a generalized adiabatic condition

$$\left|\partial_t^n(\partial_t\varphi - i\Omega^{-1}\partial_t\Omega)\right| \ll \left|\Delta\omega - i\gamma/2\right|^{n+1-k} |\Omega|^k \quad, \qquad (9)$$

where $n = 0, 1, 2,...$, $k = 0, 1, 2,..., n+1$. The adiabatic condition (9) imposes that the time-derivatives (up to given order) of the phase and the amplitude of the EMF must be much smaller than the product of the respective powers of the complex frequency detuning $(\Delta\omega - i\gamma/2)$ and the Rabi-frequency $\Omega$. This is an infinite order adiabatic condition, which unify and generalize the adiabatic condition $\left|(\Delta\omega - i\gamma)^{-1} E_o^{-1}\partial_t E_o\right| \ll 1$ [5, page 39], hereafter called standard adiabatic condition, and the Born-Fock adiabatic condition, $\left|\partial_t\Omega^{-1}\right| \ll 1$ [10, 31]. For the purpose of the present calculations, the generalized adiabatic condition of up to $n = 3$ and up to $k = 2$ is required. Although the adiabatic and the semiclassical limits are considered as equivalent [19], the adiabatic condition seems more natural from the physical point of view than the semiclassical condition ($\hbar \to 0$) and will be used here. The adiabatic condition will be applied in the following sense: neglecting $\partial_t^2 T/T$ in the brackets of Eq. (8), the explicit solution for $S(t)$ and $T(t)$ really shows that $\partial_t^2 T/T$ is much smaller than $\tilde{\Omega}'^2/4$. Then, Eq. (8) becomes

$$\partial_s^2 R + (\tilde{\Omega}'/2\partial_t S)^2 R = 0 \qquad (10)$$

To solve finally Eq. (10) we will determine the unknown variable $S(t)$ so as to makes the coefficient in front of the second term constant, *i.e.*, $\tilde{\Omega}'/2\partial_t S = const = 1$, or

$$S(t) = \int_0^t \frac{\widetilde{\Omega}'(t')}{2} dt' \tag{11}$$

This allows the solution of Eq.(10) to be taken in the form

$$R(S) = \exp(\pm iS) = \exp\left\{\pm i \int_0^t \frac{\widetilde{\Omega}'(t')}{2} dt'\right\} \tag{12}$$

Now, we are able to compose the solution of Eq. (6), which gives

$$f = T R = C_1 \sqrt{\frac{2}{\widetilde{\Omega}'}} \exp\left\{-i \int_0^t \frac{\widetilde{\Omega}'}{2} dt'\right\} + C_2 \sqrt{\frac{2}{\widetilde{\Omega}'}} \exp\left\{+i \int_0^t \frac{\widetilde{\Omega}'}{2} dt'\right\}, \tag{13}$$

where $C_1$ and $C_2$ are constants whose determination is subject of initial conditions, *e.g.*, ground state initial conditions, $g(t=0)=1$, $e(t=0)=0$. The solution (13) resemble the semiclassical form solution.

Taking into account Eq. (5), we found the following solution of the basic Eqs. (4).

$$\begin{aligned}
a_1(t) &= \sum_{j=1}^{2} C_j \sqrt{\frac{2}{\widetilde{\Omega}'}} \exp\left\{-i \int_0^t \Lambda_j dt'\right\} \\
a_2(t) &= -\frac{2}{\Omega} \sqrt{\frac{2}{\widetilde{\Omega}'}} \, [\, \sum_{j=1}^{2} C_j \widetilde{\Lambda}'_j \exp\left\{-i \int_0^t \Lambda_j dt'\right\} \,] \, \exp(i\Delta\Phi)
\end{aligned} \tag{14}$$

where

$$\Lambda_1 = (\Delta\widetilde{\omega}' + \widetilde{\Omega}')/2 \tag{15}$$
$$\Lambda_2 = (\Delta\widetilde{\omega}' - \widetilde{\Omega}')/2 \tag{16}$$
$$\widetilde{\Lambda}'_j = \Lambda_j - i(2\widetilde{\Omega}')^{-1} \partial_t \widetilde{\Omega}'. \tag{17}$$

Having solutions for the ground and excited states time-dependent amplitudes, $a_1(t)$ and $a_2(t)$ respectively, we can construct the DSs after meaningful rearrangement of the terms in the expansion of the state vector $|\Psi(\vec{r},t)\rangle$, Eq. (3). Keeping the leading non-adiabatic terms within the adiabatic condition (9), the ground $|G\rangle$ and excited $|E\rangle$ DSs are

$$\begin{aligned}
|E\rangle &= \{\, COS(\theta/2) \exp(-i\varphi) |2\rangle - SIN(\theta/2) \exp(i\omega t) |1\rangle \,\} \exp\left\{-i \int_0^t \widetilde{\omega}'_E dt'\right\} \\
|G\rangle &= \{\, SIN(\theta/2) \exp[-i(\omega t + \varphi)] |2\rangle + COS(\theta/2) |1\rangle \,\} \exp\left\{-i \int_0^t \omega_G dt'\right\}
\end{aligned} \tag{18}$$

The real and virtual components of the DSs are,:

$$|G\rangle_r = |1\rangle \exp\left\{-i\int_0^t \omega_G dt'\right\}$$

$$|G\rangle_v = |2\rangle \exp\left\{-i\left[\int_0^t (\omega_G + \omega) dt' + \varphi(t)\right]\right\}$$

$$|E\rangle_r = |2\rangle \exp\left\{-i\left[\int_0^t \widetilde{\omega}'_E dt' + \varphi(t)\right]\right\}$$

$$|E\rangle_v = |1\rangle \exp\left\{-i\int_0^t (\widetilde{\omega}'_E - \omega) dt'\right\}$$

(19)

where the field and matter frequencies where correspondingly associated so as to form the "energies" of the DSs components, Fig.1.

The quantities $\omega_G$ and $\omega_E$ are the Stark-shifted frequencies of the real ground and excited states, respectively, and $\widetilde{\omega}'_E$ will be termed instantaneous complex frequency of the real excited state. They are given by the following expressions

$$\omega_G = \omega_1 + \Lambda_2 \tag{20}$$

$$\omega_E = \omega_2 - \Lambda_2 \tag{21}$$

$$\widetilde{\omega}'_E = \omega_E - \partial_t \varphi - \gamma''/2 - i(\gamma'/2 - \Omega^{-1}\partial_t \Omega) \tag{22}$$

The quantities

$$COS(\theta/2) = (\widetilde{\Lambda}'_1 / \widetilde{\Omega}')^{1/2} \tag{23}$$

$$SIN(\theta/2) = (-\widetilde{\Lambda}'_2 / \widetilde{\Omega}')^{1/2} \tag{24}$$

are intensity-dependent amplitudes of the DS components - the partial representation of the DS components in the entire DS. They are complex quantities designated so as to underline their correspondence to the elements $\cos(\theta/2)$ and $\sin(\theta/2)$ of the unitary matrix, transforming the bare states into the DSs [6-8], as well as because they satisfy the formal condition

$$COS^2(\theta/2) + SIN^2(\theta/2) = 1 \tag{25}$$

Eliminating the dumping and field non-adiabatic factors from $COS(\theta/2)$ and $SIN(\theta/2)$ leads, of course, to $\cos(\theta/2)$ and $\sin(\theta/2)$.

The asymptotic behavior of $COS(\theta/2)$ and $SIN(\theta/2)$ with the field strength is

$$COS(\theta/2) \xrightarrow[E_o \to 0]{} 1 \tag{26}$$

$$SIN(\theta/2) \xrightarrow[E_o \to 0]{} 0 \tag{27}$$

$$COS(\theta/2) \xrightarrow[E_o \to \infty]{} \sqrt{1/2} \tag{28}$$

$$SIN(\theta/2) \xrightarrow[E_o \to \infty]{} \sqrt{1/2} \tag{29}$$

Thus, the amplitudes of the DSs (18) keep the same asymptotic behavior with the field strength as those in the perfect adiabatic case [5].

## 3. Discussion and interpretation of the results

The solutions (18), (19) represent closed form expressions for the internal dynamic of the two-level QS forced by the field and dumping non-adiabatic factors. They allow tracing explicitly the pure dynamic non-adiabatic behavior of the parameters of the DSs. The DSs (18) represents a natural generalization of the perfect adiabatic (monochromatic) field DSs solution, *e.g.*, [5-7]. The later can be reproduced from (18) eliminating all dumping and field non-adiabatic terms. The asymptotic behavior of the DSs with the field strength follows from Eqs. (26)-(29). At zero field strength, the DSs (18) reproduce the bare states. Increasing the field strength, the partial representation of the virtual components $|G\rangle_v$ and $|E\rangle_v$ increase within conditions (25) - (29). At extremely high-fields, the partial representation of the virtual and real components of given DS becomes equal. This will be called *saturation of the virtual states*. Although the present approach is nonperturbative one, the high-filed limit should be considered with caution because the RWA breaks down at high fields [13].

The DSs (18) do not obey hermitian orthogonality and do not form orthonormal basis. This is a consequence from the non-hermitian Hamiltonian (2) of the QS with dumping, considered here, as well as from the amplitude and phase variations of the EMF, which introduce additional complex value contributions to $COS(\theta/2)$ and $SIN(\theta/2)$. The DSs (18), however, generate an orthonormal non-adiabatic DSs basis neglecting the dumping and the field non-adiabatic terms in $COS(\theta/2)$ and $SIN(\theta/2)$, while keeping these terms in the exponents.

As mentioned above, the adiabatic condition (9) represents a generalization of the standard adiabatic condition and the Born-Fock adiabatic condition. They can be reproduced from (9) at $n = 0$, $k = 0$, and $n = 0$, $k = 1$, respectively, and neglecting the phase variations of the field, $\partial_t \varphi$. The field non-adiabaticity, arising from the amplitude ($\Omega^{-1}\partial_t\Omega$) and phase ($\partial_t \varphi$) variations, are expressed separately in (9). The generalized adiabatic condition (9) does not represent severe restriction for the practical application of the obtained results. It can be satisfied even for relatively fast variable fields, or short optical pulses, providing large enough frequency detuning $\Delta\omega$ from the exact resonance transitions and/or high field strength $E_o$ (Rabi frequency $\Omega$). From the other side, while the time derivatives of the phase and amplitude of the filed were retained in the initial Eq. (6), we still require the adiabatic condition (9) so as to obtain the final solutions (18), (19). That is why, Eqs. (18), (19) should be considered as weak-non-adiabatic solutions with respect to the EMF. The physical picture becomes increasingly complicated when increasing the violation of the adiabatic condition (9) because of the various ways of "penetration" of the non-adiabatic factors into the parameters of the DSs. The filed amplitude and phase non-adiabatic factors have, in general, non-separable contribution to the amplitudes of the real and virtual components of the DSs, $COS(\theta/2)$ and $SIN(\theta/2)$, respectively, as well as to the frequencies of the ground and excited DSs, $\omega_G$ and $\widetilde{\omega}'_E$.

Using Eqs. (19), one may trace out the derivation of the DSs components under the influence of the corresponding physical factors, Fig.1. The real ground state $|G\rangle_r$ is subject to complex dynamic Stark shift $\Lambda_2$, Eqs. (16), (20). The virtual ground state $|G\rangle_v$ closely follows the behavior of the real ground state $|G\rangle_r$ from which it derives. Although the virtual state takes the symmetry features of the excited state ($|G\rangle_v$ is "$|2\rangle$-type" state due to the electric-dipole coupling), its frequency/energy, subject to frequency shift $\omega$ only, follows the variations of the frequency $\omega_G$ of the real state $|G\rangle_r$ on the energy scale. In this relation, the condition (25) inspires the pictorial representation as if the real state $|G\rangle_r$, forced by the EMF, "releases" amplitude thus creating its virtual counterpart $|G\rangle_v$. This is an upward process stimulated by the field intensity, which will be called *stimulated virtual absorption* (SVA). The SVA does not coincide with the usual absorption, which takes place between two real states (in our notations $|G\rangle_r \to |E\rangle_r$), but simply leads to formation of $|G\rangle_v$. According to the adiabatic theorem of quantum mechanics, the QS will remain in given adiabatic (dressed) state, if the non-adiabatic factors acting on the QS are negligible. Transitions between different adiabatic states (in our case $|G\rangle$ and $|E\rangle$) result from non-adiabatic factors acting on the QS: non-adiabaticity of the EMF, and/or dumping [5-7]. Eqs. (16), (19) - (22) show that such a transition is accompanied by acquiring of non-adiabatic contributions from both, the field ($\partial_t \varphi$, $\Omega^{-1}\partial_t\Omega$) and the dumping factors ($\gamma''/2$, $\gamma'/2$). As Eq. (22) shows, the variation of the field phase $\partial_t \varphi$ affects the instantaneous frequency of $|E\rangle_r$, whereas the variation of the field amplitude $\Omega^{-1}\partial_t\Omega$ causes growing of the instantaneous amplitude of $|E\rangle_r$, thus populating $|E\rangle_r$ from $|G\rangle_v$. The above behavior allows us to speculate that the non-adiabatic transition $|G\rangle \to |E\rangle$ is realized by means of $|G\rangle_v \to |E\rangle_r$ transition. However, the contribution of the field phase and field amplitude non-adiabatic factors are not completely separated because the real state $|E\rangle_r$, as well as $|G\rangle_r$, are subject to complicated Stark shift, Eqs. (16), (20), (21). The dumping factors $\gamma'/2$ and $\gamma''/2$ account for the broadening and shift of the energy level, respectively. The formation of $|E\rangle_v$ is similar to that of $|G\rangle_v$, however, this time it is subject to downward process stimulated by the filed intensity, which will be called *stimulated virtual emission (SVE)*. The SVE does not coincide with the usual stimulated emission, which takes place between two real states (in our notations $|E\rangle_r \to |G\rangle_r$), but simply leads to formation of the virtual state $|E\rangle_v$. To populate $|G\rangle_r$ from $|E\rangle_v$, non-adiabatic coupling is again required. The above considerations lead to the understanding that the usual, (stimulated) absorption and the stimulated emission, must be two-step processes, mediated by formation of virtual state. If the frequency of the stimulating filed is well away from the exact resonance between the real states, *i.e.*, the adiabatic condition is well satisfied, these two steps are well separated. Then, due to the weakness of the non-adiabatic factors, the population of the real state will strongly differ from the population of the nearby-created virtual state. The above point of view is well supported by the experiment. In his elegant experiment Grischkowsky [6] has actually monitored the population of the real and the virtual states separately (in our notations, $|G\rangle_v$ and $|E\rangle_r$) thus showing the qualitative and quantitative difference in the population behavior of the real and virtual states. While close to the real state, the detuning ($\Delta\omega = 0.8 cm^{-1}$) from the exact resonance of the created virtual state in his experiment is much larger than the frequency bandwidth of the exciting laser field ($\delta\omega = 0.005 cm^{-1}$) and the effective Doppler width ($\Delta\omega_D = 0.04 cm^{-1}$). According to the frequency domain representation of the adiabatic condition ($\delta\omega \ll \Delta\omega$ [7]), the later is well satisfied and the non-adiabatic factors, while not totally absent, must be weak. As a result, the population of the real excited state $|E\rangle_r$ should be much weaker than the population of the virtual state $|G\rangle_v$, as proved by the experiment [6]. Even at well satisfied adiabatic condition ($\delta\omega$ and $\Delta\omega$ differ by more than two orders of magnitude), the transfer of population is not small effect [6]. Another important feature found is that the real and virtual state population has different time behavior [6]. If the detuning $\Delta\omega$ and the real state bandwidth $\gamma'$ are much smaller than the frequency bandwidth $\delta\omega$ of the exciting field, and, in addition, if the Stark shift is weak, *i.e.*, the non-adiabatic coupling is strong, the two-step process becomes less separable and it looks in practice as a single process. The non-adiabatic transfer of population from the virtual state to the corresponding real state can be so strong that, if the saturation of the virtual state takes place, this will result in saturation of the real state. This is what we usually call saturation of the transition, $|G\rangle_r \to |E\rangle_r$.

Tracing the phases of the states exponents in (19) inspires the understanding that one exists a causal order in the derivation of the DSs components, starting from the real component that corresponds to the initially populated bare state. Thus, starting from $|G\rangle_r$ (ground state initial condition), the phase of $|G\rangle_v$ results from that one of $|G\rangle_r$ associating the total optical phase $\omega t + \varphi$. This is in logical consent with the physical derivation of $|G\rangle_v$ from $|G\rangle_r$ by virtual absorption of a photon from the field - the SVA. The non-adiabatic factors cause population of $|E\rangle_r$ from $|G\rangle_v$. Apart from the acquired non-adiabatic phase contributions, the optical phase $\varphi(t)$ is also transferred to $|E\rangle_r$. Finally, the phase of the virtual state $|E\rangle_v$ results from that one of $|E\rangle_r$ subtracting the total optical phase. This again agrees with the physical derivation of $|E\rangle_v$ from $|E\rangle_r$ by virtual emission of one photon to the field - the SVE. Within this picture, the behavior of the phases seems physically justified although the separation of the optical phase in two parts that appear in different terms in the first of equations (18) looks at first glance unusual.

At the end, some arguments about the physical reality of the virtual components of the DSs will be considered. If the excitation (position of the created virtual state $|G\rangle_v$) is at the "red" side of the resonance ($|E\rangle_r$-state), the Stark shift is such that the real state $|E\rangle_r$ is "pushed up" while the real state $|G\rangle_r$ is "pushed down" on the energy scale, Fig.1. Opposite behavior takes place at "blue" detuning of the excitation, the real state $|E\rangle_r$ is "pushed down" while the real state $|G\rangle_r$ is "pushed up". This is a familiar behavior, well known from the theory and experiment. If we take into account the symmetry species and position of the created virtual states, one may interpreted the above behavior as repulsion of the states of same species. Note that $|E\rangle_r$ and the nearby created virtual state $|G\rangle_v$ are "$|2\rangle$-type" states, while $|G\rangle_r$ and the nearby created virtual state $|E\rangle_v$ are "$|1\rangle$-type" states, Eqs. (19). This can be considered as a specific manifestation of the *non-crossing rule* [32] for the case of QS without internal degrees of freedom, considered here. The non-crossing rule concerns usually the Born-Oppenheimer adiabatic states for QS with internal degrees of freedom, *e.g.*, molecule. Such behavior of the DSs components can be interpreted as an indication that the real DSs components "feel" the creation in their vicinity of virtual DSs component as this would be in the case of creation of any other real state. In addition, what is more important, the virtual components of the DSs have real population that can be observed experimentally [6]. This allow considering the virtual component of the DSs as a really existing state (but not an artificial mathematical construct) to same extent to what extent we consider as real any other real state.

## 4. Conclusions

Non-adiabatic dressed states for a quantum system in presence of weak field-non-adiabaticity and dumping have been derived analytically for the first time. Generalized adiabatic condition, which unifies and extends the standard and the Born-Fock adiabatic conditions, has been introduced. The non-adiabatic dressed states can generate non-adiabatic orthonormal dressed states basis. The influence of both, the field non-adiabatic factors and the dumping, on the dressed states can be traced in closed form expressions. The filed amplitude and phase non-adiabatic factors have, in general, non-separable contribution to the parameters of the DSs. Nevertheless, one may found that the field phase variations affect mainly the instantaneous frequency (energy) of the excited DSs whereas the field amplitude variations affect mainly the instantaneous amplitude of that state, thus leading to its non-adiabatic population. The interpretation of the analytic results inspires the understanding that the usual (stimulated) absorption and stimulated emission represent two-step processes, mediated by the formation of virtual state. Such understanding is in agreement with the adiabatic theorem of quantum mechanics and can be supported, analyzing existing experimental results. Evidences, showing that the virtual component of the dressed states can be considered as real physical state, are given.